\documentclass[%
reprint,
superscriptaddress,
amsmath,amssymb,
aps,
]{revtex4-2}
\usepackage{graphicx}
\usepackage{amsmath}
\usepackage{dcolumn}
\usepackage{bm}
\usepackage{hyperref}
\usepackage{xcolor}
\definecolor{Geng-color}{rgb}{0.11,0.66,0.11}

\definecolor{Ilari-color}{rgb}{0.66,0.11,0.11}

\definecolor{Mika-color}{rgb}{1,0.49,0}

\newcommand{\dd}{\mathrm{d}}

\newcommand{\NEP}{\mathrm{NEP}}
\newcommand{\FWHM}{\mathrm{FWHM}}
\newcommand{\tot}{\mathrm{tot}}

\newcommand{\John}{\mathrm{John}}
\newcommand{\Amp}{\mathrm{Amp}}
\newcommand{\eITR}{\mathrm{eITR}}
\newcommand{\opt}{\mathrm{opt}}
\newcommand{\tth}{\mathrm{th}}
\newcommand{\ph}{\mathrm{ph}}
\newcommand{\iin}{\mathrm{in}}

\begin{document}
\title{Phonon-blocked junction calorimeter}
 
	\author{Zhuoran Geng}
	\email{zhgeng@jyu.fi}
	\affiliation{%
		Nanoscience Center, Department of Physics, University of Jyv{\"a}skyl{\"a},\\ P. O. Box 35, FIN-40014 Jyv{\"a}skyl{\"a}, Finland
	}%
    \author{Joel H{\"a}tinen}
    \author{Emma Mykk{\"a}nen}
	\author{Mika Prunnila}%
	\email{mika.prunnila@vtt.fi}
	\affiliation{%
		VTT Technical Research Center of Finland Ltd.,\\ P. O. Box 1000, FIN-02044 VTT Espoo, Finland
	}%
	\author{Ilari J. Maasilta}%
	\email{maasilta@jyu.fi}
	\affiliation{%
		Nanoscience Center, Department of Physics, University of Jyv{\"a}skyl{\"a},\\ P. O. Box 35, FIN-40014 Jyv{\"a}skyl{\"a}, Finland
	}%
	\date{\today}
	
	\begin{abstract}

This study introduces the theory of a microcalorimeter based on phonon-blocked superconducting tunnel junctions, integrating on-chip electron cooling and boundary resistance phonon isolation to achieve exceptional energy resolution and rapid thermal response. A general theoretical framework is presented, along with derived approximate analytical expressions for key performance metrics, including cooling factor, thermal time constant, noise equivalent power and energy resolution. The work examines the influence of device parameters, including non-ideal effects such as subgap tunneling and heat backflow, and offers insights into optimizing the detector performance. The findings highlight the potential of the phonon-blocked junction microcalorimeters to rival or even outperform state-of-the-art technologies such as transition-edge sensors, paving the way for applications requiring precise and fast energy spectroscopy.
 \\
	\end{abstract}
	
\maketitle

\section{Introduction}
Over the past two decades, cryogenic microcalorimeters have emerged as critical tools in fields as diverse as particle physics, astrophysics, material science, and life sciences\cite{enss2005,fabjan2020a}. These detectors rely on ultra-sensitive thermometers, operating at low temperatures, to capture temperature changes caused by the energy absorption of quanta of light or particles in an absorber. The performance of these devices is typically measured by i) their energy resolution, $\Delta E$, which quantifies how precisely they can measure the energy of individual quanta, and ii) speed, limited by the time it takes for heat to dissipate from the absorber. For these calorimetric sensors, the energy resolution is inherently linked to the thermodynamic fluctuations, caused by random energy exchanges between the absorber and the heat sink, mediated typically by phonons. In a system where the absorber has a heat capacity $C_\tth$ and operates in equilibrium with a bath at a temperature $T_B$ provided by external refrigeration, the full-width half-maximum (FWHM) energy resolution corresponding to these thermal fluctuations is given by $\Delta E_\mathrm{intrinsic}=2.355\sqrt{k_BT_B^2C_\tth}$ \cite{landau1980,chui1992,enss2005}, sometimes called the thermodynamic limit \cite{nahum1995, Ullom2015}. However, this is not a hard limit, and a calorimeter can also achieve a better resolution\cite{Moseley1984,enss2005}, depending on the details of other broadband noise sources and the sensitivity of the thermometer. The theoretical FWHM energy resolution $\Delta E$ of a cryogenic microcalorimeter is still proportional to $\Delta E_\mathrm{intrinsic}$, and can therefore be  expressed as $\Delta E=\xi \Delta E_\mathrm{intrinsic}$, where the normalized, dimensionless resolution factor $\xi$ reflects the details on the sensitivity and noise characteristics of the detector, and can have values $\xi < 1$ \cite{Moseley1984,enss2005}.

Among X-ray and gamma-ray detectors, the transition-edge sensor (TES) has emerged as one of the most sensitive and widely applied microcalorimeter technologies available today\cite{irwin1995,Ullom2015,Gottardi2021}. TES uses the steep change of resistance with temperature of a superconducting film in the transition  to improve the thermometric sensitivity and thus energy resolution, achieving state-of-the-art X-ray and gamma-ray resolving powers $E/\Delta E > 1000$ \cite{Ullom2015,Gottardi2021}, which correspond in the best cases to a resolution slightly below the thermodynamic limit $\xi \leq 1$.  
Such spectroscopic capabilities are orders of magnitude better than those of conventional semiconducting energy-dispersive detectors,  setting a benchmark for other detectors to compare to. However, TES technology is not without its challenges. To set the low transition temperature $\sim 100$ mK of the film and thus the operating temperature of a TES, sophisticated normal-metal-superconductor proximity coupled bilayers typically need to be used. Therefore, the operating temperature cannot be tuned during an experiment. Such bilayers are also quite sensitive to the fabrication process details and the quality of the materials, making process control demanding. For stable operation, TES uses negative electrothermal feedback, which sets limits to the thermal relaxation time and therefore the count rate capability of the device \cite{enss2005}. This often means that TES devices need to be fabricated on suspended membranes to control phonon flow between the sensor and the thermal bath. In addition, to improve the overall count rate, large arrays often need to be used \cite{Ullom2015,Uhlig2015}.  Moreover, TES detectors require external refrigeration to cool the bath well below the TES operating temperature, to achieve optimal resolution. 

In this study, we present a general theory on an alternative low-temperature calorimeter concept, the phonon-blocked microcalorimeter, which utilizes phonon-blocking tunnel junctions\cite{mykkanen2020}, potentially addressing some of the challenges associated with TES detectors, but rivaling their performance. In this design, the radiation absorber is thermally linked to the bath only via tunnel junctions, which simultaneously operate as the electro-thermal sensing and cooling elements and as barriers to phonon heat transport. The junctions employed in this device are a pair of normal metal-insulator-superconductor (NIS) junctions connected in series (SINIS), where the normal metal island serves as both the radiation absorber and the metallic temperature sensing electrode. The N-electrode can be either a normal metal or a degenerately doped semiconductor. NIS devices, known for their use in on-chip cooling and precise thermometry in the sub-Kelvin temperature range\cite{Giazotto2006,muhonen2012}, exploit the gap in the density of states (DOS) of the superconducting electrodes as an energy filter for tunneling electrons, which with proper biasing can lead to cooling of the normal metal island. The phonon-blocking architecture significantly limits the phononic thermal conduction channel, thereby enhancing thermal isolation and improving the effectiveness of the cooling. 

Historically, NIS junction-based microbolometers, which measure radiation power rather than the energies of individual quanta, have been regarded as promising cryogenic detectors, and have been extensively studied both theoretically\cite{nahum1993,kuzmin2000,Golubev2001,Anghel2001,anghel2001b} and experimentally\cite{kuzmin2019,gordeeva2020, brien2014}. In contrast, research on NIS junction-based microcalorimeters has been relatively limited\cite{nahum1993b,nahum1995,hilton1998}, and they have been considered less effective compared to TES technology\cite{ullom2003}. However, the findings in this study suggest a more promising prospect.
By optimizing the interface properties of the junction and considering the self-cooling effect of the absorber, we demonstrate here that an optimized NIS calorimeter can achieve a normalized energy resolution of $\xi=3.3(1-\eta)$, where $\eta$ is the relative cooling factor \cite{mykkanen2020}, which has been reported to reach up to 80\% experimentally\cite{gordeeva2020}, predicting a competitive energy resolution below the thermodynamic limit.  This, combined with a fast thermal time constant, indicates that the phonon-blocked microcalorimeter could rival or even outperform the TES in energy resolution and speed.

\section{Detector model}\label{sec:model}

\begin{figure}[h]
    \centering
    \includegraphics[width=1\linewidth]{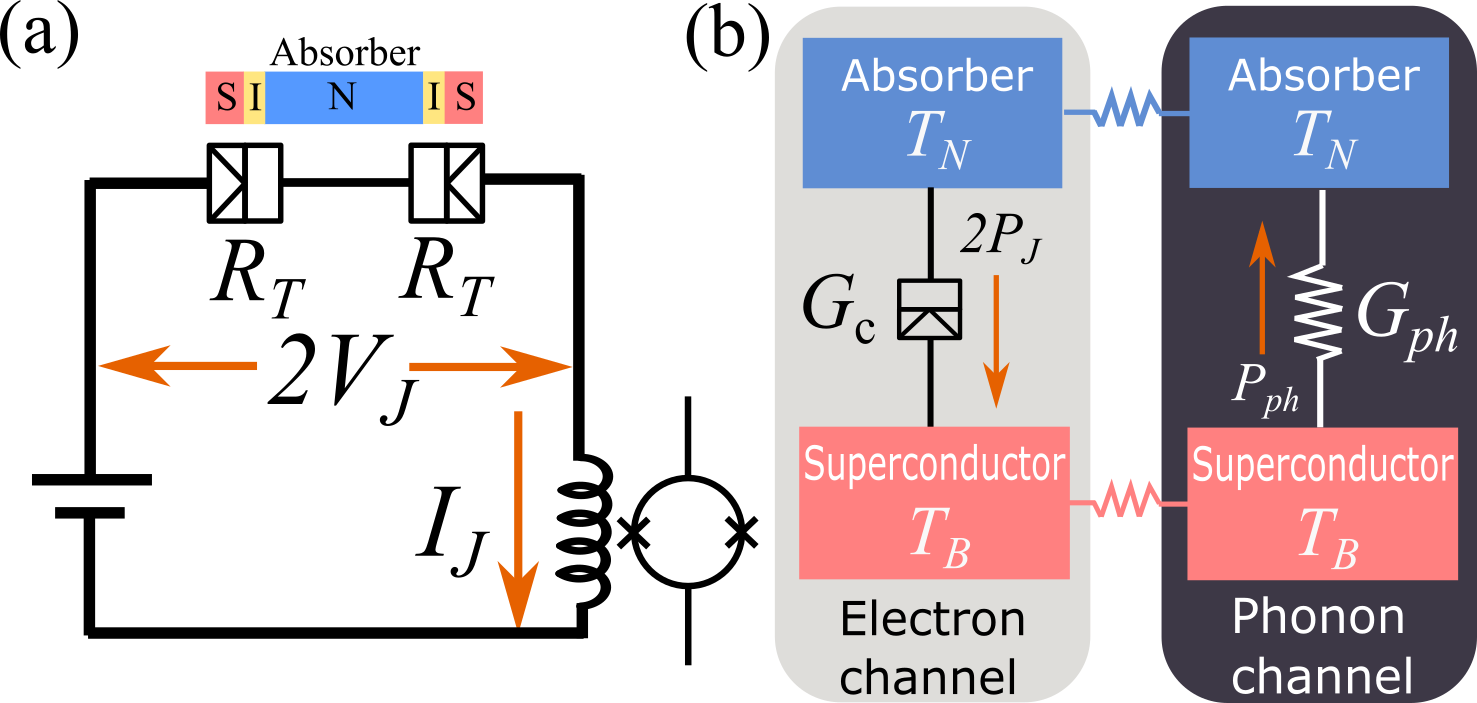}
    \caption{(a) Schematic of the the electric circuit of the system, consisting a pair of NIS junctions and readout circuit. (b) Thermal model of the system, illustrating the heat flows in the electron and phonon channels.}
    \label{fig:model_circuit}
\end{figure}

To detect the energies of radiation quanta, a calorimetric device typically comprises several key components, each designed to fulfill specific functions\cite{Ullom2015}. These include an absorber to capture the incident photon and convert its energy to heat; a sensing element to transduce this thermal excitation into a measurable electrical signal; and a thermal resistance to control heat flow between the sensor and the bath, ensuring proper thermal isolation.

In our model, a large normal metal island functions as the absorber, while two identical NIS tunnel junctions, acting as the temperature sensors and coolers, are created by adding superconducting contacts to the common normal metal island, separated by thin insulating tunnel barriers. Figure \ref{fig:model_circuit}(a) presents a simple electrical circuit model of the device and its readout, where each voltage biased NIS junction has a tunneling resistance $R_T$ and a voltage drop $V_J$. The resulting temperature dependent current signal, $I_J$, is read by a superconducting quantum interference device (SQUID) through an input coil connected in series with the tunnel junctions.

The heat flow between the sensor and the bath occurs through both electron and phonon channels, as illustrated in Fig. \ref{fig:model_circuit}(b). A crucial aspect of our model is that phonons can travel to the bath only through the tunnel junctions, which provide significant thermal resistance for the phonon channel\cite{mykkanen2020,kemppinen2021,hatinen2023} due to interfacial thermal resistance\cite{Swartz1989} (acoustic mismatch). The resulting thermal phonon power flow into the absorber is then governed by the equation
\begin{equation}
    P_\ph = \frac{1}{4a_\eITR}(T_B^4-T_N^4),
\end{equation}
where $a_\eITR$ is the effective interfacial thermal resistance coefficient, and $T_B$ and $T_N$ are the temperatures of the bath and the absorber, respectively. Our model
assumes that heat flow is limited by the junctions and not by the electron-phonon coupling in the relatively large volume absorber, causing the electron and phonon systems in the absorber to remain at the same temperature $T_N$. In addition, we also assume that the temperature of the superconducting electrode equilibrates with that of the thermal bath $T_B$. This allows us to study the theoretical performance limits with the most efficient cooling performance of the junctions. In practice, some engineering is typically required to avoid the overheating of the superconductor, for example by coupling the superconducting electrodes to additional thermalizing normal metal regions (quasiparticle traps) \cite{pekola2000, muhonen2012, Giazotto2006}.  Consequently, we define the heat currents in the electron  channel as $2P_J$, the cooling power generated by the the two tunnel junctions (positive for heat flow out of the absorber), and in the phonon channel as $P_\ph$, the heat current associated with the phonon flow (positive for heat flow into the absorber).

When an incident power $P_\iin$ is absorbed by the detector, the temperature $T_N$ rises in accordance with the heat balance equation derived from the thermal model in Fig.\ref{fig:model_circuit}(b):
\begin{equation}\label{eq:thermal_eq}
     C_\tth\frac{\dd}{\dd t} T_N+2P_J-P_\ph = P_\iin+2\beta(P_J+I_JV_J).
\end{equation}
In this expression, the heat capacity $C_\tth=\gamma V T_N$ of the absorber is assumed to be dominated by the electron system, where $\gamma$ is the Sommerfeld coefficient and $V$ the absorber volume. The last term with the factor $2\beta$ describes a heat backflow non-ideality: it accounts for the portion $0 \le \beta < 1$ of heat dissipated by quasiparticles in the superconducting electrodes that returns to the absorber\cite{fisher1999}, adding to the input signal and phonon transport heat. We also assume that the resistivity of the normal metal absorber is low enough that direct Joule heating of it is negligible \cite{RNnote}.

The electric current $I_J$ and cooling power $2P_J$ generated by the pair of NIS junctions are given by\cite{jochum1998,Chaudhuri2012}
\begin{align}
     I_J(V_J,T_N,T_B) =& \frac{1}{eR_T}\int^{\infty}_{-\infty}\dd  \epsilon\  N_S\big(\epsilon,T_B\big) \nonumber  \\
     &\times \big[
            f(\epsilon,T_B)-f(\epsilon+eV_J,T_N)
            \big], \label{eq:IJ_expr} \\
     P_J(V_J,T_N,T_B) =& \frac{1}{e^2R_T}\int^{\infty}_{-\infty}\dd  \epsilon\ \big(\epsilon+eV_J\big)N_S(\epsilon,T_B)  \nonumber\\
     &\times \big[
            f(\epsilon+eV_J,T_N)-f(\epsilon,T_B)
            \big].\label{eq:PJ_expr}
\end{align}
Here $e$ is the absolute value of the electron charge, $f(\epsilon,T)$ is the Fermi-Dirac distribution, and $N_S(\epsilon,T_B)$ is the broadened superconductor density of states
\begin{equation*}
    N_S(\epsilon,T_B)=\bigg|\mathrm{Re}\bigg[
    \frac{\epsilon/\Delta_0+i\Gamma}{\sqrt{(\epsilon/\Delta_0+i\Gamma)^2-\Delta^2(T_B)/\Delta_0^2}}
    \bigg]\bigg|,
\end{equation*}
where $\Delta(T_B)$ and $\Delta_0$ denote the superconducting gap at temperature $T_B$ and at the zero temperature limit, respectively, and $\Gamma$ is the dimensionless phenomenological Dynes parameter that influences predominantly the subgap tunneling behavior\cite{dynes1978,pekola2010}. 

At low temperatures ($k_BT_N, k_BT_S\ll \Delta_0$), the optimal bias voltage $V_b = V_\mathrm{opt}$ at each junction for maximizing cooling power $P_J$ can be approximated by $eV_\mathrm{opt}=\Delta_0-0.66k_BT_N$\cite{Anghel2001,Giazotto2006,muhonen2012}. Under this optimal bias, the junction current and cooling power simplify to\cite{mykkanen2020,Giazotto2006}:
\begin{align}
    I_J(V_\opt)\approx &0.48\frac{\sqrt{k_BT_N\Delta_0}}{eR_T}, \label{eq:IJ_approx} \\
     P_J(V_\opt) \approx & \frac{\Delta_0^2}{e^2R_T}\Bigg[
        0.59\bigg(\frac{k_B T_N}{\Delta_0}\bigg)^{\frac{3}{2}}
        -\sqrt{\frac{2\pi}{\beta_B}}e^{-\beta_B}\Bigg] \nonumber\\
        &-\frac{V_\opt^2}{2R_T}\Gamma,  \label{eq:PJ_approx}
\end{align}
where $\beta_B=\Delta_0/k_BT_B$ is introduced for clarity. In addition, the heat current ratio flowing between the two channels can also be approximated in the low-temperature limit and for $\Gamma = 0$ by $P_\ph/P_J(V_\opt)\approx 0.42\rho[T_B^4/(T_N^{1.5}T_c^{2.5})-(T_N/T_c)^{2.5}]$ , where $T_c$ is the critical temperature of the superconductor, and where we have introduced
\begin{equation}
    \rho = \frac{R_T}{a_\eITR}\frac{e^2T_c^2}{k_B^2\sqrt{\Delta_0/k_BT_c}}
\end{equation}
as a dimensionless parameter characterizing the thermal resistivity ratio between the electron and phonon channels. It describes an interface property of the tunnel junction independent of the junction area $A$, as both $R_T$ and $a_\eITR \propto 1/A$, with values $\rho < 1$ ($\rho > 1$) corresponding to electron (phonon) channel dominating the heat flow.

\begin{figure}
    \centering
    \includegraphics[width=1\linewidth]{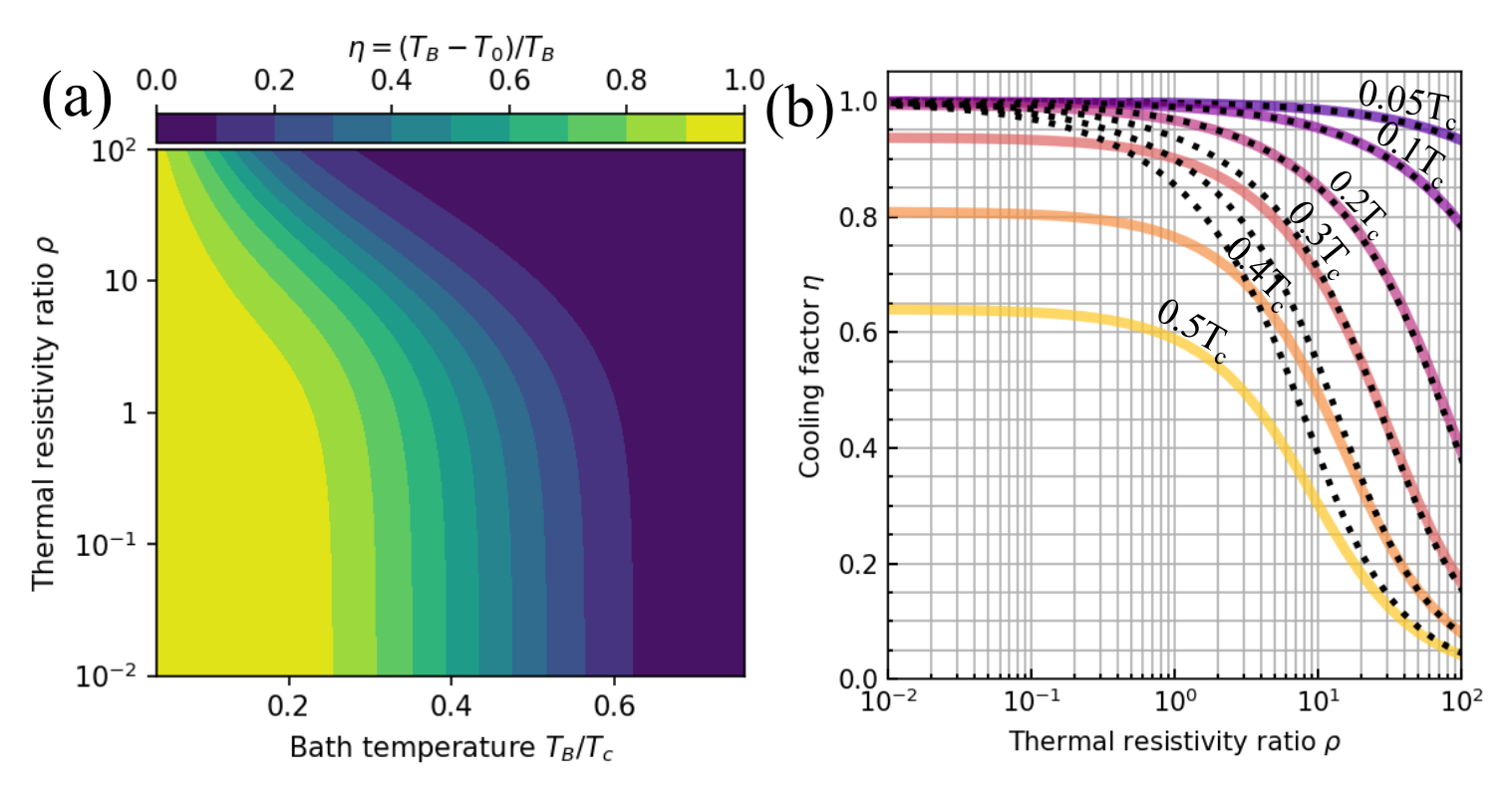}
    \caption{(a) The relative cooling factor $\eta$ as a function of the normalized bath temperature $T_B/T_c$, and thermal resistivity ratio $\rho$. (b) Comparison of $\eta$ between the numerical result (colored solid lines) and the analytical approximation, Eq. \eqref{eq:TN_approx} (black dotted lines) as a function of $\rho$.}
    \label{fig:fig2_cooling}
\end{figure}

In the quiescent state, where $P_\iin=0$, the heat balance equation \eqref{eq:thermal_eq} becomes $2P_J(V_b,T_N,T_B)=P_\ph(T_N,T_B)$, assuming an idealized detector with $\beta=0$. In such a state, the absorber reaches a steady state temperature $T_{N,0}\equiv T_0 < T_B$, which can be approximated in the low temperature limit and for $\Gamma = 0$ by solving the equation
\begin{equation}\label{eq:TN_approx}
    \bigg(\frac{T_B}{T_c}\bigg)^4-\bigg(\frac{T_0}{T_c}\bigg)^4=\frac{4.72}{\rho}\bigg(\frac{T_0}{T_c}\bigg)^{1.5}
    .
\end{equation}

To provide a dimensionless measure of cooling performance, we introduce the relative cooling factor defined as\cite{mykkanen2020}
\begin{equation}
    \eta = \frac{T_B-T_0}{T_B}.
\end{equation}
Fig. \ref{fig:fig2_cooling}(a) shows the numerically optimized cooling factor $\eta$ in the steady state for an ideal device with $\beta = 0$, $\Gamma = 0$, where Eq. \eqref{eq:PJ_expr} was used in the heat balance equation instead of solving the approximate expression, Eq. \eqref{eq:TN_approx}.  We see that $\eta$ increases as both the bath temperature $T_B$ and the resistivity ratio $\rho$ decrease. One notable feature is that $\eta$ approaches a high limiting value at low $\rho<1$, shown more clearly in panel (b), in which $\eta$ is plotted as a function of $\rho$ for different bath temperatures (colored solid lines). Panel (b) also compares the numerical results (colored solid lines) with the analytical approximation from Eq. \eqref{eq:TN_approx} (black dotted lines), showing excellent agreement between the two for bath temperatures below $0.3T_c$.

\section{Calorimetric response}
When a photon is absorbed by the detector, its energy is instantly converted into energetic excitations within a small volume of the absorber. These excitations (electrons, phonons) undergo diffusion and scattering, quickly distributing their energy as heat, forming a thermal quasi-equilibrium. The combined thermal equilibration and diffusion time of the charge carriers within the metal absorber is typically of the order of or less than a microsecond, which is much shorter than the thermal time constant $\tau_\tth$, the time it takes for heat to dissipate from the absorber to the thermal bath. Due to this disparity in timescales, it is typically assumed that the absorber experiences an initial step-like temperature rise, given by $\Delta T = E / C_\tth$\cite{Moseley1984,enss2005}, where $E$ is the energy of the photon and $C_\tth$ is the heat capacity of the absorber. This rapid temperature rise, followed by its gradual decay, is measured by a temperature-sensitive element, which in this case are the NIS tunnel junctions. By tracking the current signal $I_J(t)$ over time, the energy of the photon can be accurately determined.

The detector's current responsivity $S_I$, which defines the small signal power-to-current transfer function, plays a key role in characterizing the conversion between the input heat and the measurable electrical signal. When the input power variation $\Delta P_\iin$ is small enough that the temperature change $\Delta T /T_N \ll 1$, the responsivity $S_I = \Delta I_J / \Delta P_\iin$ can be derived using a first-order linearization of the thermal and electrical balance equations of the detector. To express this relationship, we define $T_0$, $I_0$, and $V_0$ as the absorber temperature, junction current, and single junction voltage in the quiescent state, respectively. By denoting the small deviations of those variables $\Delta T_N = T_N - T_0$, $\Delta I_J = I_J - I_0$, and $\Delta V_J = V_J - V_0$, we can write
\begin{equation}\label{eq:linearized_balance_eqs}
    \begin{aligned}
    \Delta P_\iin =& (1+i\omega\tau_\tth)G_\tth\Delta T_N-2\beta V_0\Delta I_J\\
    & +\frac{V_0}{R_J}b_0\Delta V_J  +\delta P,\\
    2\Delta V_J=&-i\omega L\Delta I_J+\delta V, \\
    \Delta I_J =&\alpha_T\frac{I_0}{T_0}\Delta T_N+\frac{1}{R_J}\Delta V_J +\delta I,
\end{aligned}
\end{equation}
where the first equation follows from the thermal balance Eq. \eqref{eq:thermal_eq}, the second from the electrical circuit equation (Fig. \ref{fig:model_circuit}), where $L$ is the input inductance of the SQUID, and the third from a first-order Taylor expansion. Here, the equations are expressed in the frequency domain, with $\omega$ representing the angular frequency of the signal. The terms $\delta P$, $\delta I$, and $\delta V$ refer to small perturbations due to intrinsic fluctuations (noise).

The linear coefficients appearing in Eqs.\eqref{eq:linearized_balance_eqs} are critical for understanding the detector's thermal and electrical response. These coefficients include the dynamic thermal conductance of the phonon channel, $G_\ph=|\partial P_\ph/\partial T_N|_{T_N = T_0} = T_0^3 / a_\mathrm{eITR}$, and the total dynamic thermal conductance in the electron channel, $G_c = 2(\partial P_J / \partial T_N)|_{V_0}$. With these, we define $G_\tth = G_\ph + (1 - \beta) G_c$ as the total thermal conductance, including the fraction from the backflow, and $\tau_\tth = C_\tth / G_\tth$ as the thermal time constant of the system. In addition,  the dynamic resistance of a single junction is defined as $R_J = (\partial V_J / \partial I_J)|_{T_0}$, while the dimensionless sensitivities to temperature and voltage of a junction current and power are given by $\alpha_T = (\partial \ln I_J / \partial \ln T_N)|_{V_0}$ and $\alpha_V = (\partial \ln P_J / \partial \ln V_J)|_{T_0}$, respectively. $b_0 = 2\left[\alpha_V - \beta(\alpha_V + I_0V_0 / P_{J0})\right] \times \left(P_{J0} R_J / V_0^2\right)$ is a dimensionless constant which simplifies calculations of the current response. 

Equations \eqref{eq:linearized_balance_eqs} demonstrate the interconnected nature of the thermal and electrical behavior of the detector through the state variables $\Delta P$, $\Delta I$, and $\Delta V$. In the ideal case where $\beta=0$ and for the optimal bias (for which $\alpha_V = 0$), the constant $b_0 = 0$, and the equations simplify to $\Delta P_\iin = (1+i\omega\tau_\tth)G_\tth\Delta T_N+\delta P$ and $\Delta I_J =(\alpha_T I_0/T_0)\Delta T_N++i\omega\tau_e\Delta I_J+\delta I$, where $\tau_e = L/(2 R_J)$ is the electrical time constant, eliminating feedback from the output signal to the temperature change and simplifying the analysis.

With this simplification, the frequency-dependent responsivity can be derived as 
\begin{equation}\label{eq:SI_expr}
    S_I(\omega)=\frac{\mathcal{L}_0}{V_{0}(1+i\omega\tau_\tth)(1+i\omega\tau_e)} \approx \frac{\mathcal{L}_0}{V_{0}(1+i\omega\tau_\tth)},
\end{equation}
where $\mathcal{L}_0=\alpha_TI_0V_0/G_\tth T_0$ represents a dimensionless factor analogous to the zero-frequency loop gain commonly used in calorimetric analysis\cite{enss2005,Giazotto2006}, and the last approximation is good in the typical case, where $\tau_e \ll \tau_\tth$. 

\begin{figure}[ht]
    \centering
    \includegraphics[width=1\linewidth]{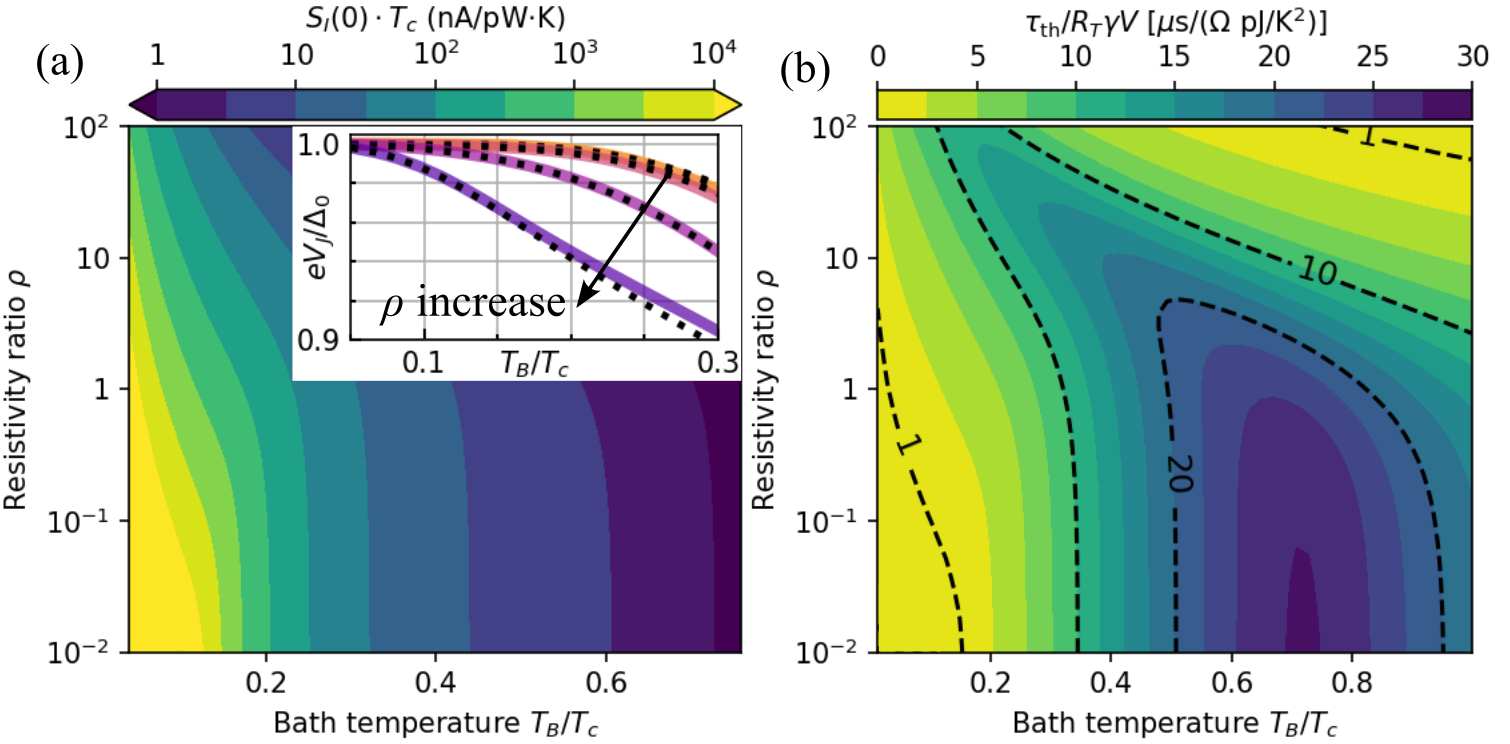}
    \caption{(a) Optimal zero-frequency responsivity $S_I(0)$ as a function of the normalized bath temperature $T_B/T_c$ and resistivity ratio $\rho$. The inset compares the numerically optimized bias voltage  with the approximate $V_\opt$ at three different $\rho = 0.1,1,10$ from top to bottom. (b) The thermal time constant $\tau_\tth$ at optimal bias, scaled by $\gamma V R_T$, as a function of $T_B/T_c$ and $\rho$.}
    \label{fig:responsivity}
\end{figure}

Equation \eqref{eq:SI_expr} indicates that the responsivity has a cutoff frequency at $\omega_c = \tau_\tth^{-1}$ and remains constant at low frequencies $\omega \tau_\tth \ll 1$, defining the zero-frequency responsivity $S_I(0) = \mathcal{L}_0/V_{0}$. Figure \ref{fig:responsivity}(a) illustrates the numerical results for $S_I(0)$ as a function of the bath temperature $T_B$ and the resistivity ratio $\rho$ for an ideal device and for the optimal bias that minimizes $T_0$ at each point $(T_B,\rho)$, showing that decreasing $T_B$ and $\rho$ increases $S_I(0)$. The bias voltage corresponding to the main plot at three fixed $\rho$ is plotted in the inset of panel (a)  by the colored solid lines, which align closely with the approximate optimal bias $eV_\mathrm{opt} = \Delta_0-0.66k_BT_0$, represented by the black dotted line, at $T_B<0.3T_c$.

This alignment allows us to derive an approximate expression for the optimal $S_I$. At $V_\opt$, the total thermal conductance can be derived using Eqs. \eqref{eq:IJ_approx} and \eqref{eq:PJ_approx} to be $G_\tth \approx T_0^3 / a_\mathrm{eITR} + 1.77 k_B \sqrt{k_B T_0 \Delta_0} / e^2 R_T$, while the temperature sensitivity is approximately $\alpha_T \approx 1$ for all $T_0 \ll \Delta_0/k_B$. Consequently, the low-frequency responsivity can be approximated as
\begin{equation}\label{eq:SI_approx}
    S_I(0)\approx0.48\frac{e}{k_BT_c}\Bigg[
        1.77\frac{T_0}{T_c}+\rho\bigg(\frac{T_0}{T_c}\bigg)^{3.5}
    \Bigg]^{-1}.
\end{equation}

In this expression, the second term corresponding to the phonon channel diminishes rapidly with decreasing temperature, simplifying Eq. \eqref{eq:SI_approx} to approximately $S_I(0) \approx 0.27(e/k_B)/T_0 = 3.15 / T_0[K]$ (nA/pW) in the low-temperature limit, depending only on the inverse of the electron temperature of the absorber, and being independent of its volume. This result is lower than that reported in previous research\cite{Golubev2001} by a factor of two, owing to the double junction configuration used in this study.

The thermal time constant, $\tau_\tth$, characterizes the evolution of the absorber temperature over time, following in the small signal limit the expression $\Delta T_N(t) = (E / C_\tth) \exp(-t / \tau_\tth)$ for NIS calorimeters without feedback\cite{nahum1993b}. At the optimal bias, $\tau_\tth$ can be approximated at the quiescent temperature $T_0 \ll \Delta_0/k_B$ as
\begin{equation} \label{eq:tau_thermal}
\begin{aligned}
\tau_\tth= C_\tth/G_\tth \approx \gamma V\bigg(\frac{T_0^2}{a_\eITR}+1.77\frac{k_B^2}{e^2R_T}\sqrt{\frac{\Delta_0}{k_BT_0}}\bigg)^{-1}\\
     = \gamma VR_T \frac{e^2}{k_B^2}\sqrt{\frac{k_BT_c}{\Delta_0}}\Bigg[ \rho\left(\frac{T_0}{T_c}\right)^2+1.77\left(\frac{T_c}{T_0}\right)^{\frac{1}{2}}\Bigg]^{-1}.
     \end{aligned}
\end{equation}

This expression indicates that $\tau_\tth$ always decreases at low temperatures, and also at higher temperatures for $\rho \gtrsim 1$,  creating in that case a maximum with a pivot point at $T_0/T_c = (0.44/\rho)^{0.4}$. Figure \ref{fig:responsivity}(b) illustrates the numerically computed thermal time constant at optimal bias conditions as a function of $T_B$ and $\rho$, in scaled units $\tau_\tth/(\gamma VR_T)$, showing a peaked distribution as a function of bath temperature. We see that the detector is fastest either at low temperatures and where the electron channel dominates thermal transport ($\rho < 1$), or alternatively at high temperatures and where the phonon channel dominates ($\rho \gg 1$). However, only the first option offers good responsivity (Fig. \ref{fig:responsivity} (a)). In the low-temperature limit, where the thermal time constant is primarily governed by the electrical channel, $\tau_\tth\propto \gamma VR_T\sqrt{T_0}$. This suggests that operating at lower temperatures enhances both the sensitivity and the speed of the detector. This behavior is a notable characteristic that sets the phonon-blocked microcalorimeter apart from other more conventional cryogenic microcalorimeters. In contrast, a TES detector slows down  at lower operating temperatures\cite{enss2005}. For example, using a realistic value $C_\tth=0.8$ pJ/K at 0.1 K, the thermal time constant is $\tau_\tth/R_T < 50\ \mu$s/$\Omega$ when $\rho$ is low, outperforming state-of-the-art TES devices with similar heat capacities\cite{ullom2005, lee2015, miniussi2018, bandler2019, sakai2023}.

Additionally, it is worth noting that for an ideal detector, the zero-frequency responsivity $S_I(0)$ is governed by the resistivity ratio $\rho$, and is therefore independent of the dimensions of the tunnel junctions. In contrast, the detector speed, determined by the thermal time constant, scales with the junction area $A$  as $1/\tau_\tth \propto A$. Notably, in the low-temperature limit, Eq.\eqref{eq:SI_approx} reduces to its first term, eliminating explicit dependence on $R_T$. This suggests that for applications requiring high signal bandwidth, using large-area tunnel junctions is beneficial, as that can increase bandwidth without compromising responsivity.

\section{Noise and energy resolution}

The FWHM energy resolution, $\Delta E$, is the key measure of the performance of a microcalorimeter in detecting the energy of individual quanta. While the introduction addressed the fact that the resolution is intrinsically linked to thermodynamic fluctuations of energy between the absorber and the bath, as expressed by $\Delta E_\mathrm{intrinsic} = 2.35\sqrt{k_B T_B^2 C_\tth}$, all major noise sources must be considered in practical devices in more detail for accurate performance analysis. For our device, the first important contribution is from the power fluctuations due to thermal phonons traveling in and out of the absorber, often called phonon noise. Another unavoidable major contributor is shot noise of the junctions, which arises from fluctuations in the number of electrons passing through the voltage-biased tunnel junction, generating noise in both electrical and heat currents\cite{Blanter2000,Anghel2001,Golubev2001}. Furthermore, amplifier noise from the readout electronics and Johnson noise from the resistance of the absorber can also contribute to the overall noise budget. As a result, only signal power exceeding the combined noise power from all these sources will carry useful information about the absorbed particle.

To accurately account for these various noise sources, $\Delta E$ can be estimated by examining the total noise power spectral density in the frequency domain. For stationary noise, the measurement in each angular frequency interval, $\dd \omega$, is statistically independent of others, and the FWHM energy resolution, assuming optimal filtering and an instantaneous absorption event\cite{enss2005},  is given by\cite{enss2005,Moseley1984}
\begin{equation}\label{eq:dE_def}
    \Delta E = 2.35\bigg[
        \int_0^\infty \frac{2}{\pi} \frac{1}{\NEP_\tot^2(\omega)}\dd \omega
    \bigg]^{-\frac{1}{2}},
\end{equation}
where $\NEP_\tot(\omega)$ represents the total noise equivalent power, which defines the signal power required in a 1-Hz bandwidth to match the noise.

The total NEP is derived by summing all independently contributing noise sources
\begin{equation}\label{eq:NEPcomponent}
\NEP_\tot^2=\NEP_\ph^2+2\NEP_J^2+\NEP_\Amp^2+\NEP_\John^2,
\end{equation}
where each term represents a specific source of noise. These contributions can be evaluated from the small power variation, $\langle\Delta P^2\rangle$, induced by the fluctuations applied to the perturbation terms in Eqs.\eqref{eq:linearized_balance_eqs}. Specifically, $\NEP_\ph$ corresponds to thermodynamic phonon power fluctuations, $\delta P_\ph$; $\NEP_J$ arises from shot noise at each junction (giving the factor of two for the total), incorporating both power fluctuations due to the tunneling electrons, $\delta P_J$, and current fluctuations, $\delta I_J$; $\NEP_\Amp$ stems from readout noise in terms of equivalent input current fluctuations, $\delta I_\Amp$; and $\NEP_\John$, which represents voltage fluctuations, $\delta V_\John$, generated by the resistance of the absorber, and is considered negligible compared to all other sources in this study.

As a result, the total NEP can be expressed as
\begin{equation}\label{eq:tot_NEP}
\begin{aligned}
    \NEP_\tot^2(\omega)&=\langle\delta P_\ph^2\rangle 
    + \frac{\langle\delta I_\Amp^2\rangle}{|S_I(\omega)|^2} \\
    &+ 2\Bigg[\langle\delta P_J^2\rangle + \frac{\langle\delta I_J^2\rangle}{|S_I(\omega)|^2} + 2\mathrm{Re}\bigg(\frac{1}{S_I(\omega)}\bigg)\langle\delta I_J\delta P_J\rangle
    \Bigg],
\end{aligned}
\end{equation}
where we assume no correlations between phonon noise, junction noise, and amplifier noise, but include the unavoidable cross-correlation between the junction electrical and heat current fluctuations\cite{Golubev2001} (each tunneling event carries both charge and energy at the same time). It is also worth noting that the cross-correlation term $\langle\delta I_J\delta P_J\rangle$ is negatively valued under optimal bias conditions, leading to a reduction in the total NEP.

At quiescent temperature $T_0$, for the contribution from phonon noise, we use the expression
\begin{equation}\label{eq:SPph_eq}
\begin{aligned}
    \langle\delta P_\ph^2\rangle &= \frac{2k_B}{a_\eITR}(T_0^5+T_B^5)\\
&= 2k_B(G_{\ph}T_0^2+G_{\textrm{bath}}T_B^2),
\end{aligned}
    \end{equation}
where $G_{\textrm{bath}} = \partial P_\ph/\partial T_B = T_B^3/a_\eITR$ and $G_\ph = |\partial P_\ph/\partial T_N|_{T_N = T_0} = T_0^3/a_\eITR$. It is exact in the two opposite limits of ballistic or small transmission, as the correction term is proportional to $\mathcal{T}(1-\mathcal{T})$, where $\mathcal{T}$ is the phonon transmission probability\cite{krive2001}. In our case, we have $\mathcal{T} \ll 1$ and therefore ignore the small correction.  

The noise spectral density components of the junction, which include both current and power fluctuations as well as their cross-correlation, are given \cite{Golubev2001} as follows:
\begin{equation}\label{eq:junction_noise}
    \begin{aligned}
     \langle\delta I_J^2\rangle =& \frac{2}{R_T}\int^{\infty}_{-\infty}\dd  \epsilon\ N_S(\epsilon,T_B)\big(
            f_N+f_S-2f_Nf_S
            \big), \\  
     \langle\delta P_J^2\rangle =& \frac{2}{e^2R_T}\int^{\infty}_{-\infty}\dd  \epsilon\ \big(\epsilon+eV_J\big)^2N_S(\epsilon,T_B)  \\
     &\times \big(
            f_N+f_S-2f_Nf_S
            \big), \\
     \langle\delta I_J\delta P_J\rangle =& \frac{2}{eR_T}\int^{\infty}_{-\infty}\dd  \epsilon\ \big(\epsilon+eV_J\big)N_S(\epsilon,T_B)  \\
     &\times \big(
            f_N+f_S-2f_Nf_S
            \big),       
    \end{aligned}
\end{equation}
where the Fermi-Dirac distributions of the absorber $f_N = f(\epsilon + eV_J,T_0)$ and the superconducting electrodes $f_S = f(\epsilon, T_B)$ are evaluated at temperatures $T_0$ and $T_B$, respectively.

At low temperatures and around the optimal voltage bias, simple approximations can be used for the tunnel current noise and cross-correlation noise in Eqs.\eqref{eq:junction_noise}. These are expressed as $\langle\delta I_J^2\rangle \approx 2e|I_J|$ and $\langle\delta I_J P_J\rangle \approx -2eP_J$\cite{Golubev2001}. In contrast, the low-temperature analytical expression for junction heat current noise at optimal bias can be derived to consist of three key terms:
\begin{equation}\label{eq:SPJ_approx}
\begin{aligned}
    \langle\delta P_J^2\rangle &\approx 4\sqrt{2\pi}\frac{\Delta_0^3}{e^2R_T}\bigg(\frac{k_BT_B}{\Delta_0}\bigg)^{\frac{1}{2}}e^{-\Delta_0/k_BT_B} \\
    &+2.05\frac{\Delta_0^3}{e^2R_T}\bigg(\frac{k_BT_0}{\Delta_0}\bigg)^{\frac{5}{2}} +\frac{2}{3}\frac{(eV_\opt)^3}{e^2R_T}\Gamma.
\end{aligned}
\end{equation}
Here, the first two terms originate from thermal quasiparticle fluctuations resulting from the finite temperatures of the superconductor and the normal metal absorber, while the third term accounts for the shot noise from the subgap tunneling current, characterized by $\Gamma$.

\begin{figure*}[t]
    \centering
    \includegraphics[width=0.7\linewidth]{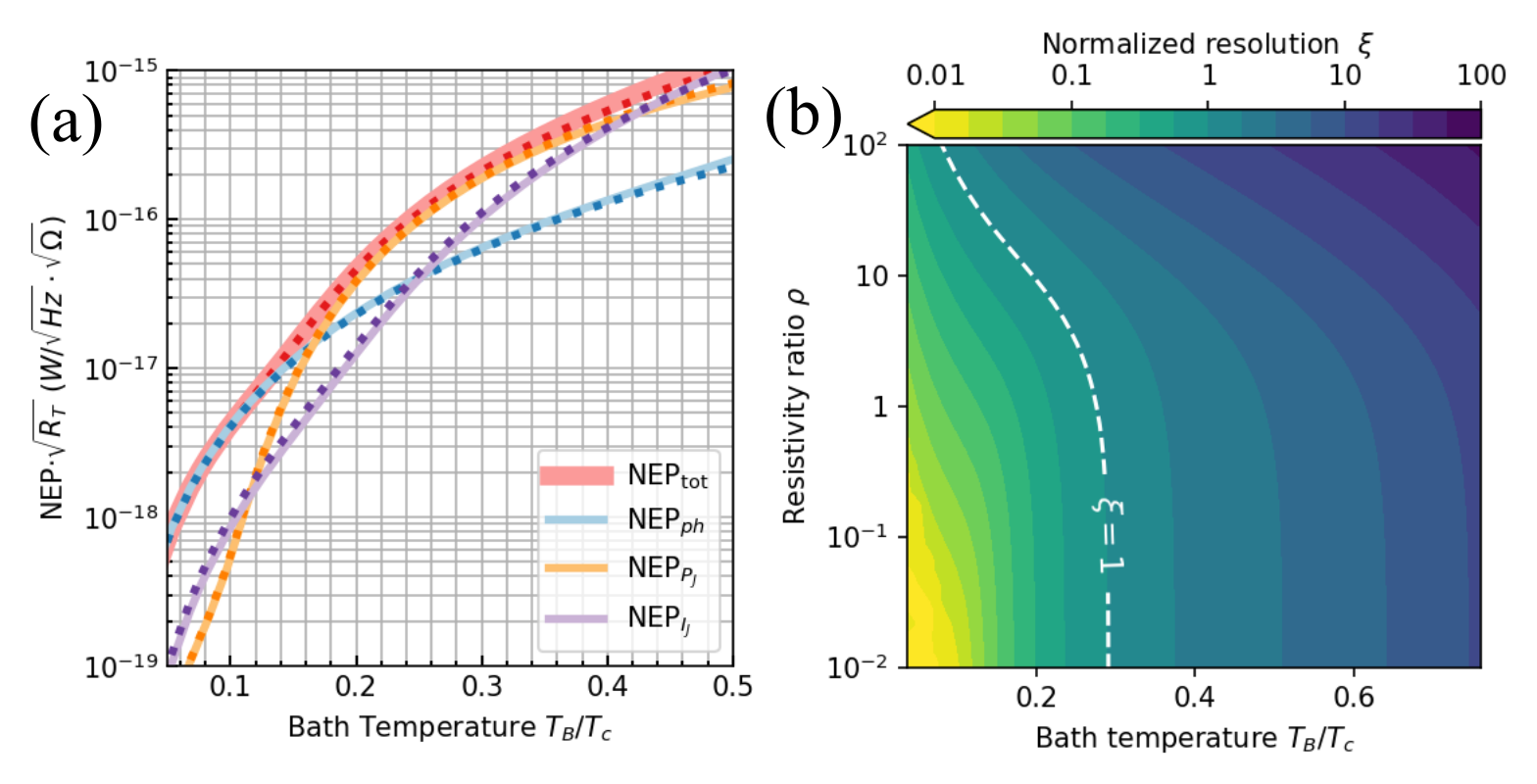}
    \caption{(a) Numerically optimized total low-frequency NEP at optimal bias and its decomposition into various contributions in Eq.\eqref{eq:NEPcomponent} (solid colored lines), scaled as $\NEP\cdot\sqrt{R_T}$, as a function of $T_B/T_c$ for an ideal detector with $\rho=1$. Numerical results are also compared with the analytical approximations (dotted colored lines). (b) Numerically calculated normalized energy resolution $\xi$ as a function of $T_B/T_c$ and $\rho$.}
    \label{fig:NEP_and_dE}
\end{figure*}

Substituting  the analytical expressions from Eqs.\eqref{eq:IJ_approx},\eqref{eq:PJ_approx},\eqref{eq:SI_approx}, and \eqref{eq:SPJ_approx} into Eq.\eqref{eq:tot_NEP} and performing algebraic simplifications, the low temperature total NEP of the detector at $V_\opt$ becomes
\begin{equation}\label{eq:NEP_approx}
    \begin{aligned}
        \NEP^2_\tot(\omega) &\approx 4k_BT_B^2G_\tth(1-\eta)^2\bigg[
        \frac{(1-\kappa)}{2}  \left(1+\frac{1}{(1-\eta)^5}\right)\\
        &+2.8\frac{\kappa\beta_B^2}{(1-\eta)^{\frac{5}{2}}}e^{-\beta_B} + 0.19\kappa\Gamma\beta_N^{\frac{5}{2}} + 0.58\kappa \\
        &+3.7\frac{1}{\kappa\alpha_T^2}(1+\omega^2\tau_\tth^2) -2.46\frac{1}{\alpha_T} +2.08\frac{\Gamma}{\alpha_T}\beta_N^{\frac{3}{2}}\\
        &+1.9\frac{R_T}{\Delta_0}\frac{1}{\kappa\alpha_T^2}\sqrt{\frac{\beta_B}{1-\eta}}(1+\omega^2\tau_\tth^2)\langle\delta I_\Amp^2\rangle
        \bigg],
    \end{aligned}
\end{equation}
where $\kappa=G_c/G_\tth=1/[1+0.56\rho(T_0/T_c)^{2.5}]$ is the relative strength of the thermal conductance in the electron channel at optimal bias, and $\beta_B=\Delta_0/k_BT_B$ and $\beta_N=eV_\opt/k_BT_0$ are introduced for clarity. The first row represents the contribution from phonon noise, which becomes insignificant if the thermal conductance from phonons, $G_\ph$, is much smaller than that from the junction heat current, $G_c$, i.e. when $\kappa \rightarrow 1$. The second row captures the noise power from the junction heat current, with the first term exponentially suppressed at low  $T_B$. The third row represents the noise generated by the electrical current, where the first term takes into account the cutoff frequency limiting the high-frequency responsivity, whereas the second, negative and third, positive terms reflect the noise correlation between the electrical and the heat currents in the junction. Finally, the last term accounts for the readout noise, which will be neglected in the discussion of an ideal detector.

Figure \ref{fig:NEP_and_dE}(a) illustrates as an example the low-frequency ($\omega\tau_\tth \ll 1$) NEP of an ideal detector ($\Gamma = 0, \beta = 0)$ with $\rho=1$, operating at its optimal bias point\cite{NEPopt}. The red solid line represents the numerically calculated total NEP, which decreases with bath temperature and is inversely proportional to $\sqrt{R_T}$. This behavior indicates that more resistive junctions generate less noise, a trend that contrasts with the detector's thermal time constant $\tau_\tth$, which increases with $R_T$ (see Eq. \eqref{eq:tau_thermal}). The red dotted line corresponds to the approximate expression of Eq.\eqref{eq:NEP_approx}, showing excellent agreement with the numerical results up to 0.5$T_c$. This consistency provides valuable insight into the noise characteristics of the detector.

To further understand the noise behavior, panel (a) also breaks down the contributions from individual noise components, computed using both numerical methods (solid colored lines) and the analytical approximations of Eq. \eqref{eq:NEP_approx} (dotted colored lines). At higher bath temperatures $T_B/T_c>0.44$, the thermal conductance ratio $\kappa$ becomes much smaller than one, making the electric current noise, NEP$^2_{I_J}=2\langle\delta I_J^2\rangle/|S_I(\omega)|^2$, the dominant source. As the temperature decreases to intermediate level, $\kappa$ increases, leading to a more significant contribution from the junction's heat current noise, NEP$^2_{P_J}=2\langle\delta P_J^2\rangle$. In the low-temperature limit $T_B/T_c <0.16$, the phonon noise, NEP$^2_{\ph}=\langle\delta P_\ph^2\rangle$, begins to dominate. This aligns with Eq.\eqref{eq:NEP_approx}, as the cooling factor $\eta$ in an ideal detector starts to approach $1$, amplifying the first term in the noise expression. Note that, although we have included it in the total NEP curves, we haven't separately plotted the correlation term proportional to $\langle\delta I_J\delta P_J\rangle$, as it is negative at optimal bias and also because it is not dominant at any temperature. 

This analysis also reveals a fundamental low-temperature limit for the NEP of an ideal detector, where the heat flow occurs exclusively through the electrical channel of the junction ($\kappa=1$). In this regime, the NEP approaches
\begin{equation}
\NEP_\tot(0) = 2.7(1-\eta)\sqrt{k_B T_B^2 G_\tth},
\end{equation}
setting the ultimate noise floor achievable by the detector. Note that the above achieves a value below the noise floor of the phonon noise limited equilibrium case $\NEP_{\ph,eq}(0) = 2\sqrt{k_B T_B^2 G_\tth}$ for $\eta > 0.26$. 

The FWHM energy resolution $\Delta E$ can be obtained in a straightforward manner using Eq.\eqref{eq:dE_def}. At the optimal bias, it reads as
\begin{equation}
    \Delta E = \xi\times2.35\sqrt{k_BT_B^2C_\tth},
\end{equation}
where $\xi$ can be approximated at low temperatures ($k_BT_N, k_BT_S\ll \Delta_0$) by
\begin{equation}\label{eq:dE_approx}
    \begin{aligned}
        \xi =& 2\frac{1-\eta}{\sqrt{\alpha_T}}\Bigg[
        1.8\frac{1-\kappa}{\kappa}\left(1+\frac{1}{(1-\eta)^5}\right) \\
        &+\big(0.7\Gamma+10.5\beta_B^{-\frac{1}{2}}e^{-\beta_B}\big)\beta_N^{\frac{5}{2}} + 2.1 \\
        &+ \frac{1}{\kappa^2\alpha_T^2}\big(13.7-9.1\kappa\alpha_T + 7.7\kappa\alpha_T\Gamma\beta_N^{\frac{3}{2}}\big)
        \Bigg]^\frac{1}{4}.
    \end{aligned}
\end{equation}

Figure \ref{fig:NEP_and_dE}(b) presents the numerically calculated normalized resolution $\xi$ as a function of normalized bath temperature ($T_B/T_c$) and resistivity ratio $\rho$ for an ideal detector ($\Gamma =0$, $\beta = 0$) biased at $V_\opt$. We see that $\xi$ decreases with lower $T_B$ and smaller $\rho$, eventually approaching saturation when $\rho$ is small. Notably, $\xi$ can drop well below 1, indicating that the energy resolution can surpass the intrinsic thermodynamic limit. The threshold for $\xi<1$ is denoted by the white dashed line in the figure, occurring at approximately $0.3T_c$ for $\rho<1$.

At low temperatures, a fundamental resolution limit can be derived under the same assumptions as used for the limiting NEP, by setting $\kappa = 1$. This yields a resolution of $\Delta E\approx3.2(1-\eta)\times \Delta E_\mathrm{intrinsic}$. A similar factor was derived by Nahum and Martinis\cite{nahum1993b,nahum1995}, but their model did not account for the cooling effect ($\eta = 0$). This result suggests that achieving a cooling factor $\eta>70\%$ enables phonon-blocked microcalorimeters to surpass the intrinsic thermodynamic limit.

In comparison, the optimal resolution for an ideal TES detector is given by $\xi\approx2.4/\sqrt{\alpha_\mathrm{TES}}$\cite{irwin1995,enss2005}, where $\alpha_\mathrm{TES}$ can exceed 100, predicting low values $\xi \ll 1$. To match the best theoretical TES performance, NIS junctions would require an on-chip cooling factor of approximately 90\%. However, the highest resolution real, non-ideal TES X-ray detectors have not attained factors below $\xi \approx 0.4$ \cite{miniussi2018}. Encouragingly, recent experiments have also shown that NIS tunnel junctions can achieve cooling factors greater than 80\% at bath temperatures of 200-300 mK\cite{gordeeva2020}, and cooling of bulk silicon plates with a factor of 40\% at $T_B \approx 200$ mK has also been demonstrated\cite{mykkanen2020}.

\section{Non-ideal device}\label{sec:dicsussion}
In previous sections, the general theory for a phonon-blocked microcalorimeter was established, and numerical results were presented for an ideal device. In this section, the focus moves to estimating the performance of a more practical device using a set of realistic parameters. This approach provides insight into the expected performance of a real-world detector and serves as a baseline to showcase the impact of non-ideal parameters on the detector performance.

\begin{table}
    \centering
    \begin{tabular}{cl}
    \hline
    \hline
       Zero $T$ superconducting gap $\Delta_0$ & 0.2 meV \\
       Tunneling resistance $R_T$ & 2.5 $\Omega$ \\
       Dynes parameter $\Gamma$ & $10^{-3}$ \\
       Heat return coefficient $\beta$ & 0\\
       Interfacial thermal resistance coeff. $a_\mathrm{eITR}$ & 0.10 K$^4$/nW \\
       Absorber Sommerfeld coefficient$\times$volume $\gamma V$ & $2$ pJ/K$^2$ \\
       Input coil inductance $L$ & 1.8 $\mu$H \\
       SQUID readout noise $\langle\delta I_\Amp^2\rangle$ & 60 fA$/\sqrt{Hz}$ \\
    \hline
    \hline
    \end{tabular}
    \caption{Key parameters of the proposed device of Fig. \ref{fig:real_device}.}
    \label{tab:parameters}
\end{table}

The key parameters of the device investigated in this section are summarized in Table \ref{tab:parameters}. 
The tunnel junctions are modeled with a superconducting gap $\Delta_0=0.2$ meV, \textit{e.g.} aluminum, and an area of $80$ $\mu m^2$/ junction. Each junction has a tunneling resistance of $R_T=2.5\ \Omega$ by assuming a specific tunneling resistance of $200\ \Omega\,\mu m^2$, which is well within the experimentally demonstrated range for Al junctions\cite{kemppinen2021,mykkanen2020,nguyen_2013}. The two phenomenological parameters, the Dynes parameter $\Gamma$ and the heat return coefficient $\beta$, are here first set to 10$^{-3}$ and 0\cite{gunnarsson2015,kemppinen2021}, respectively. We will address the impact of a range of both these parameters more thoroughly below. 

The phonon channel is assumed to be limited by interfacial thermal resistance of the junctions, and the specific resistance is set to $15.4$ K$^4\mu m^2$/nW, corresponding to an experimental value\cite{hatinen2023}, leading to an effective interfacial thermal resistance coefficient of $a_\eITR=0.096$ K$^4$/nW. These parameter values correspond to a thermal resistivity ratio $\rho = 4.6$, assuming the BCS relation between $T_c$ and $\Delta_0$, $\Delta_0 = 1.76 k_BT_c$. 

The heat capacity of the absorber is assumed to be dominated by the electron system within the temperature range of interest, with $\gamma V=2$ pJ/K$^2$, a value comparable to the highest resolution practical TES absorbers for X-ray detection \cite{lee2015,miniussi2018}, resulting in a temperature increase of less than 5\% at 0.1 K for an absorbed 6 keV photon, and keeping the device therefore well in the linear response regime. We stress that this choice of heat capacity is not optimized as other system parameters (required stopping power, X-ray energy, spectral bandwidth, etc.) typically constrain the absorber properties. It also remains to be analyzed how far the phonon-blocked calorimeter can be pushed towards the non-linear regime, i.e. what its dynamic range is.    

Finally, we also take into account the added noise of a practical SQUID amplifier to be used in the readout, with the values for the input coil inductance and amplifier input current noise listed in Table I based on commercially available products\cite{wolf2017}. For the chosen device parameters, we found that the SQUID noise gives an insignificant contribution to the total detector noise budget. 

\begin{figure*}
    \centering
    \includegraphics[width=0.7\linewidth]{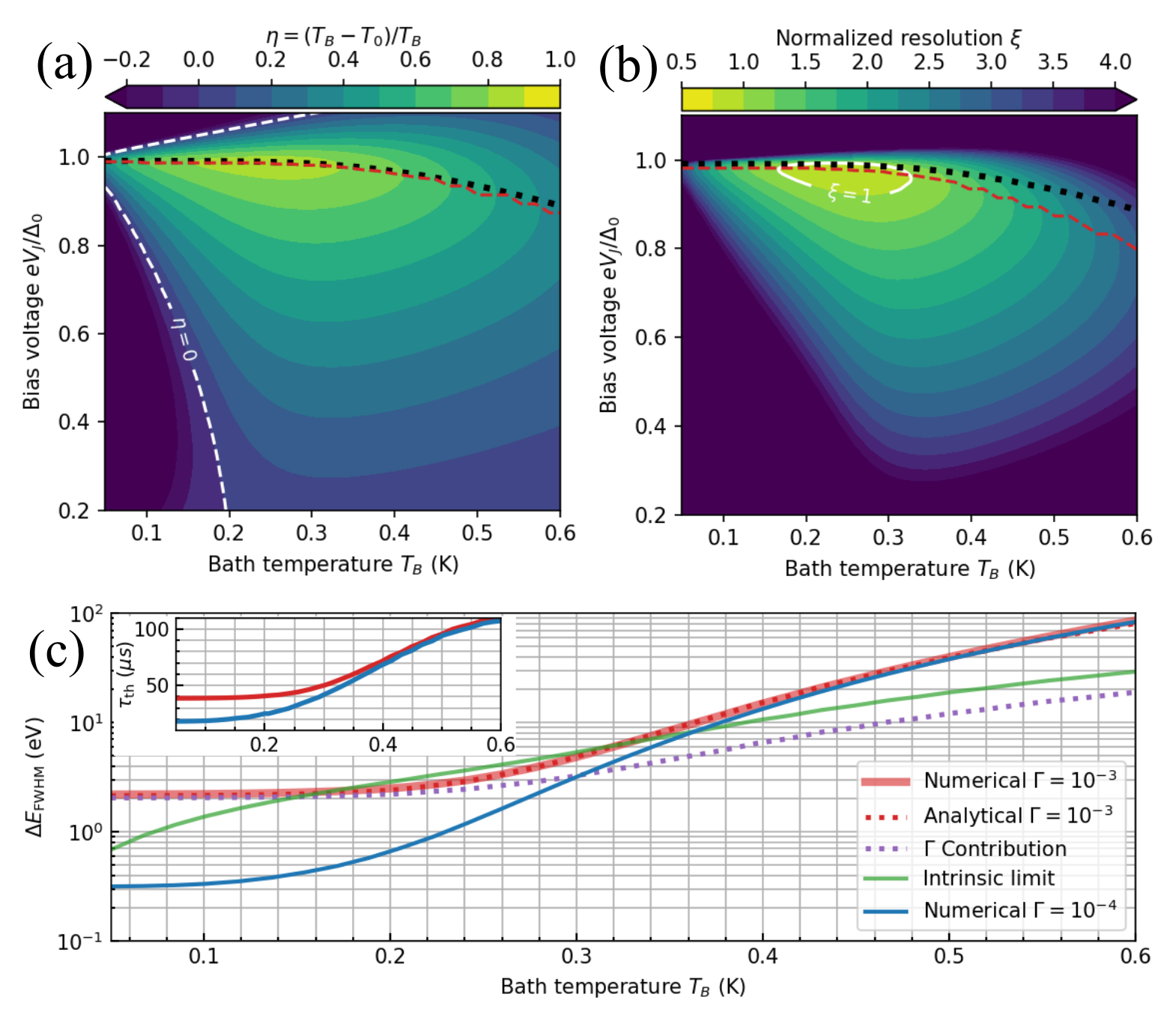}
    \caption{(a)  Cooling factor $\eta$ and (b) normalized resolution $\xi$ as a function of bath temperature and bias voltage of a practical detector with $\Gamma = 10^{-3}$. Dashed red line shows the numerical optimal bias voltage, dashed black line the approximation $V_{opt}$. (c) Comparison of the energy resolution results for $\Gamma = 10^{-3}$ (red solid line) and $\Gamma = 10^{-4}$ (blue solid line) as a function of bath temperature. Green solid line = $\Delta E_{\textrm{intrinsic}}$. Inset: thermal time constant as a function of $T_B$. }
    \label{fig:real_device}
\end{figure*}

Figure \ref{fig:real_device} provides a detailed look at the numerically computed performance of such a practical microcalorimeter, showcasing how realistic device parameters affect both the cooling factor $\eta$ and the FWHM energy resolution $\Delta E$. In panel (a), the cooling factor $\eta$ is plotted as a function of the bath temperature $T_B$ and the bias voltage $V_J$. The red dashed line indicates the bias voltage that maximizes cooling, closely following the approximate optimal bias $V_\opt = \Delta_0 - 0.66k_BT_0$ (black dotted line). Moving away from this optimal point, the cooling effectiveness diminishes, and at a certain range of voltages, it even switches to heating, with the boundary where $\eta=0$ shown by the white dashed lines. Unlike an ideal detector, where $\eta$ increases monotonously with lowering temperature, the practical device achieves a maximum of 85\% around 0.25 K, close to that reported in experiment\cite{gordeeva2020}. This limitation is attributed to subgap tunneling effects, characterized by $\Gamma = 10^{-3}$, which dominate the detector performance at low temperatures. Subgap tunneling smears the singularity of the superconducting density of states by allowing states to appear within the gap\cite{dynes1978}, reducing the efficiency of hot electron filtering and causing excess Joule heating of the normal metal\cite{Pekola2004}. This effect is responsible for the additional voltage-dependent heating term in Eq.\eqref{eq:PJ_approx}, which, being independent of bath temperature, reduces cooling factor at low temperatures $T < 0.2$ K.

Panel (b) presents the numerically computed normalized energy resolution $\xi$ as a function of the bath temperature and the bias voltage. The red dashed line indicates the voltage that minimizes $\xi$, located slightly below $V_\opt$ (black dashed line). This shift occurs because subgap tunneling introduces an additional resolution penalty that depends on $V_J$, slightly lowering the optimal bias to minimize $\xi$. Notably, a region where $\xi<1$ appears around a bath temperature of 0.25 K, enclosed by the white solid line, indicating that the device can achieve an energy resolution below the thermodynamic limit, with a minimum $\xi$ of 0.8. Another remark is that $\xi$ at the optimal bias is not a particularly sensitive function of $T_B$: even at $T= 0.5$ K it has only increased by a factor of two from the thermodynamic limit.

Finally, panel (c) displays the numerically calculated FWHM energy resolution at the optimal bias as a function of the bath temperature, shown as the red solid line, with the red dotted line showing the analytical approximation of Eq.\eqref{eq:dE_approx}, which matches very closely with the numerical results in the whole bath temperature range shown. This alignment then provides a basis for understanding how each parameter influences the energy resolution. At bath temperatures below 0.15 K, the energy resolution saturates around $\Delta E \approx 2$ eV, largely limited by the junction power fluctuation noise due to subgap tunneling (the term proportional to $\Gamma\beta_N^{\frac{5}{2}}$ in Eq. \eqref{eq:dE_approx}), depicted by the violet dotted line. This limitation is also confirmed by comparing it with the numerical result of a detector with the same $\rho$ but a lower $\Gamma = 10^{-4}$ (blue solid line), revealing a degradation in the resolution of the $\Gamma = 10^{-3}$ device below $T_B = 0.4$ K. We see that the device with $\Gamma = 10^{-4}$ is predicted to achieve an exceptionally good $\Delta E \approx 0.3$ eV at $T_B < 0.1$ K well below the values for state-of-the-art X-ray TES detectors\cite{lee2015,miniussi2018}. Nevertheless, even the more conservative device with the higher $\Gamma = 10^{-3}$ achieves a resolution below the thermodynamic limit $\Delta E_\mathrm{intrinsic}$ (green solid line) within the temperature range of 0.15 K to 0.3 K, illustrating its robustness to non-idealities of that magnitude. Additionally, the inset in panel (c) highlights the relatively fast thermal response, with a time constant of approximately $\tau_\tth =$ 50 $\mu$s below 0.3 K, underscoring that the detector is also suitable for applications requiring high X-ray photon count rates. 

As demonstrated in the preceding analysis, subgap tunneling substantially reduces detector performance. To optimize the device, reducing this non-ideal leakage current is essential. In practical applications, the Dynes parameter $\Gamma$ is typically considered to be a phenomenological parameter that can reflect several different microscopic non-idealities, such as photon-assisted tunneling\cite{pekola2010} or two-particle Andreev current\cite{faivre2015}, causing $\Gamma$ to depend, for example, on the transparency of the junction barrier and the electromagnetic environment of the junctions. For aluminum-based cooler junctions, typical values for $\Gamma$ are around $10^{-3}-10^{-4}$ \cite{gunnarsson2015,mykkanen2020,oneil2012,nguyen2014}, though values as low as $< 5 \times 10^{-7}$ have been achieved with advanced filtering and shielding \cite{pekola2010,saira2012} for junctions with high $R_T \sim 1 M\Omega$. Additionally, recent studies suggest that hybrid superconducting/ferromagnetic structures may effectively suppress the Andreev current, possibly lowering $\Gamma$ and thus enhancing junction cooling\cite{gordeeva2020, giazotto2002, ozaeta2012}.

\begin{figure}
    \centering
    \includegraphics[width=\linewidth]{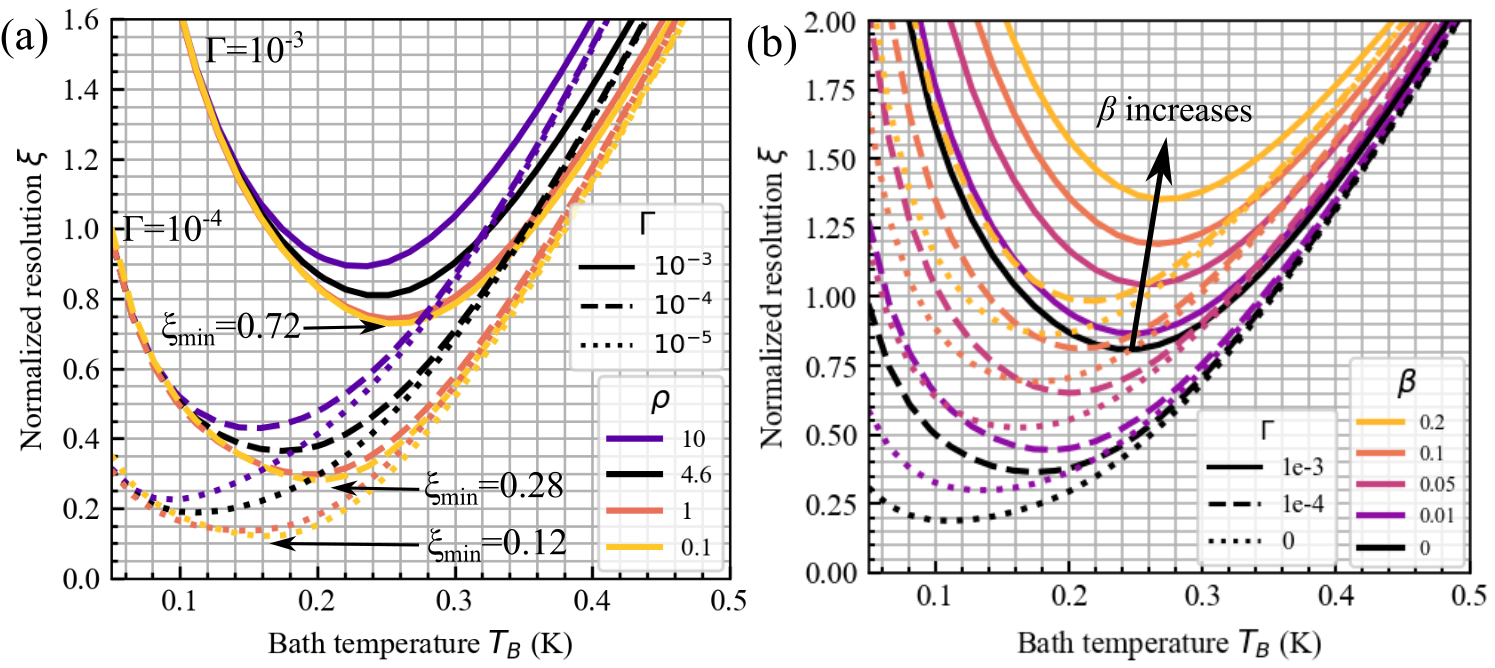}
    \caption{(a) Normalized energy resolution $\xi$ as a function of bath temperature for different values of the Dynes parameter $\Gamma = 10^{-3}$, $10^{-4}$, $10^{-5}$ , indicated by solid, dashed and dotted lines, respectively, and for different $\rho$, indicated by different colors. (b) $\xi$ versus heat backflow coefficient $\beta$ (colored) at varying values of $\Gamma$, with $\rho = 4.6$.}
    \label{fig:fig6}
\end{figure}

Figure \ref{fig:fig6}(a) illustrates how reducing $\Gamma$ improves $\xi$ by comparing three values of $\Gamma$ with all other parameters held constant, as shown with the black lines for the device with the parameters from Table I ($\rho = 4.6$). Lowering $\Gamma$ from $10^{-3}$ (black solid line) to $10^{-4}$ (black dashed line) reduces the minimum $\xi$ from 0.8 to 0.36, achieving an energy resolution $\Delta E_\FWHM=0.5$~eV at a bath temperature of 170 mK. This strong improvement in energy resolution can be seen more clearly in Fig. \ref{fig:real_device}(c) from the blue solid line. It saturates at a lower temperature $T_B<0.1$ K with a lower absolute resolution about 0.3 eV, significantly lower than the reported TES detectors. In addition, it also further improves the speed of the detector, reaching $20\ \mu$s at low temperature. Further decreasing of the subgap leakage (i.e., $\Gamma = 10^{-5}$ depicted as black dotted line in Fig. \ref{fig:fig6}(a)) improves $\xi$ even further to 0.12 and allows operation at lower optimal bath temperatures.


Beyond reducing $\Gamma$, the performance of the detector can be further enhanced by optimizing the resistivity ratio $\rho$. 
For the device demonstrated in Figure \ref{fig:fig6} (a), the decrease in $\rho$ generally improves $\xi$, as shown by the colored lines: at $\rho=0.1$, $\xi$ reaches the minimum of 0.72 ($\Gamma = 10^{-3}$, solid yellow line) and 0.28 ($\Gamma = 10^{-4}$, dashed yellow line), approaching the theoretical limits for ideal TES detectors. Further reduction below $\rho=0.1$ yields diminishing returns, consistent with the saturation trends in Figure \ref{fig:NEP_and_dE}(b).

Lastly, we briefly address the influence of a finite heat backflow from the superconducting electrode to the normal metal, characterized by the phenomenological parameter $\beta$. A nonzero $\beta$ reduces the net cooling power proportionally, as the backflow power scales with both junction cooling and Joule heating power as $2\beta(P_J + I_JV_J)$. Figure \ref{fig:fig6}(b) illustrates how increasing $\beta$ degrades the normalized energy resolution. For instance, with $\Gamma = 10^{-3}$ (solid colored lines), values of $\beta > 0.05$ push the resolution above the intrinsic thermodynamic limit ($\xi>1$). For lower values of $\Gamma$, the tolerance to backflow increases; for example, at $\Gamma = 10^{-4}$ (dashed colored lines), $\beta$ must exceed 0.2 to compromise the resolution above the thermodynamic limit. This again emphasizes the importance of minimizing the subgap tunneling. In practice, backflow can be mitigated by incorporating quasiparticle traps, such as an additional normal metal layer on the superconductor electrode \cite{pekola2000, oneil2012, nguyen_2013}. For such devices, measured value of $\beta < 0.02$ has been reported\cite{oneil2012}.

\section{Conclusions}
In this work, we have presented a general theory of a self-cooled microcalorimeter based on phonon-blocked tunnel junctions, demonstrating its potential as a highly sensitive and fast cryogenic detector. In this detector concept, tunnel junctions simultaneously serve three purposes: as sensors, as cooling elements, and as barriers to phonon thermal conductance. A key contribution of this study is the derivation of approximate analytical expressions for critical performance metrics, such as the responsivity, the noise equivalent power and the energy resolution. These expressions not only provide insights into the fundamental performance of an ideal detector but also serve as valuable tools for guiding the optimization of non-ideal devices, by accounting for practical limitations such as the finite sub-gap leakage current, the readout noise, and a significant phonon contribution of the total heat conductance. 

Our analysis reveals that, under ideal conditions, the energy resolution of the phonon-blocked junction microcalorimeter can achieve a value of approximately $3.3(1-\eta)\Delta E_\mathrm{intrinsic}$, rivaling TES detectors if appreciable cooling factors $\eta$ are achieved. Notably, we identified that the key detector performance metrics, such as the cooling factor $\eta$, the responsivity $S_I$, and the normalized energy resolution $\xi$, are largely independent of the junction area. Instead, they are governed by the dimensionless parameter $\rho \propto R_T/a_\eITR$, where $R_T$ is the normal-state electrical tunneling resistance and $a_\eITR$ interfacial phonon thermal resistance coefficient. This independence, coupled with the finding that the thermal time constant $\tau_\tth \propto R_T\sqrt{T_0}$ ($T_0$ being the quiescent temperature), underscores the potential to achieve both high resolution and fast thermal response by employing large-area junctions at low temperatures.

The study also highlights the challenges posed by non-idealities, particularly subgap tunneling, which can significantly degrade the performance. However, our practical device example demonstrates that the detector can achieve energy resolution below the intrinsic thermal fluctuation noise limit ($\xi<1$) with parameters already experimentally demonstrated. With a further optimization of subgap tunneling, an area where recent experiments using hybrid superconducting/ferromagnetic structures have already shown promising results, phonon-blocked junction microcalorimeters have the potential to outperform even the theoretical performance limits of TES detectors.

In conclusion, the phonon-blocked junction microcalorimeter represents a promising alternative to the state-of-the-art cryogenic calorimeter technology, offering exceptional energy resolution and fast thermal response. Its ability to achieve on-chip cooling minimizes the need for complex cryogenic setups, making it ideal for applications requiring precise energy-resolving photon detection at high count rates. With further advancements in reducing subgap tunneling and optimizing interface properties, such a detector has the potential to set a new benchmark in cryogenic microcalorimetry and expand the possibilities for future scientific and technological advancements.

\begin{acknowledgments}

This research was supported at JyU by the Research Council of Finland projects number 341823 and 359240 (the Finnish Quantum Flagship, University of Jyv{\"a}skyl{\"a}). VTT acknowledges for the financial support the European Union’s Horizon RIA and EIC programme grant No. 101113086 SoCool and Research Council of Finland through project No. 350667 Femto, No. 336817 QTF Centre of Excellence project and No. 374172 QMAT Centre of Excellence project, and Technology Industries of Finland Centennial Foundation.

\end{acknowledgments}

\bibliography{reference}
\end{document}